\magnification=\magstep1
\tracingpages=1
\input epsf

\let\mc=\ninerm


\def\Clint{1}
\def\Conway{2}
\def\Sbase{3}
\def\SP{4}
\def\Dsyn{5}
\def\Dsemi{6}
\def\Ehrlich{7}
\def\Floyd{8}
\def\Ki{9}
\def\KO{10}
\def\Ksp{11}
\def\Kiv{12}
\def\Spiders{13}
\def\KR{14}
\def\PR{15}
\def\Ruskey{16}
\def\Squire{17}
\def\Steiner{18}
\def\VM{19}
\def\WD{20}

\def\\#1{\hbox{\it#1\/\kern.05em}}
{\catcode`\;=\active
\gdef\beginprogram{$$\catcode`\;=\active\bf
                    \def;{{\rm\char`\;}\ }
                    \halign\bgroup\hskip3em##\hfil\cr}
\gdef\beginwideprogram{$$\catcode`\;=\active\bf
                    \def;{{\rm\char`\;}\ }
                    \halign\bgroup\hskip1em##\hfil\cr}}
\def\label#1: {\llap{\rm#1: }}
\def\endprogram{\egroup$$}
\def\scope{{\rm scope}}
\def\section #1. #2. {\bigskip\noindent{\bf#1.\enspace#2.\enspace}}
\def\dash---{\thinspace---\hskip.16667em\relax}
\mathcode`\@="8000 {\catcode`\@=\active \gdef@{\mkern1mu}}

\centerline{\bf Efficient Coroutine Generation of Constrained Gray Sequences}
\smallskip
\centerline{Donald E. Knuth and Frank Ruskey}
\centerline{(dedicated to the memory of Ole-Johan Dahl)}

\bigskip\begingroup\narrower
\noindent{\bf Abstract.} We study an interesting family of
cooperating coroutines, which is able to generate all
patterns of bits that satisfy certain fairly general
ordering constraints, changing only one bit at a time. (More
precisely, the directed graph of constraints is required to
be cycle-free when it is regarded as an undirected graph.)
If the coroutines are implemented carefully, they yield an
algorithm that needs only a bounded amount of computation
per bit change, thereby solving an open problem in the field
of combinatorial pattern generation.
\par\endgroup

\bigskip\noindent
Much has been written about the transformation of procedures
from recursive to iterative form, but little is known about
the more general problem of transforming {\it coroutines\/} into
equivalent programs that avoid unnecessary overhead. The
present paper attempts to take a step in that direction by
focusing on a reasonably simple yet nontrivial family of
cooperating coroutines for which significant improvements in
efficiency are possible when appropriate transformations are
applied. The authors hope that this example will inspire
other researchers to develop and explore the potentially
rich field of coroutine transformation.

Coroutines, originally introduced by M. E. Conway [\Conway], are analogous to
subroutines, but they are symmetrical with respect
to caller and callee: When coroutine~$A$ invokes coroutine~$B$, the action
of~$A$ is temporarily suspended and the action of~$B$ resumes where $B$ had
most recently left off. Coroutines arise naturally in producer/consumer
situations or multipass processes, analogous to the ``pipes'' of {\mc UNIX},
when each coroutine transforms an input stream to an output stream; a
sequence of such processes can be controlled in such a way that their
intermediate data files need not be written in memory.
(See, for example, Section 1.4.2 of~[\Ki].)

The programming language {\mc SIMULA~67} [\Sbase] introduced support for
coroutines in terms of fundamental operations named
{\bf call}, {\bf detach}, and
{\bf resume}. Arne Wang and Ole-Johan Dahl subsequently discovered [\WD] that
an extremely simple computational model is able to accommodate these
primitive operations. Dahl published several examples to demonstrate their
usefulness in his chapter of the book {\sl Structured Programming\/} [\SP];
then M.~Clint~[\Clint] and O.-J. Dahl~[\Dsemi] began to develop theoretical
tools for formal proofs of coroutine correctness.

Another significant early work appeared in R. W. Floyd's general top-down
parsing algorithm for context-free languages [\Floyd], an algorithm that
involved ``imaginary men who are assumed to automatically appear when hired,
disappear when fired, remember the names of their subordinates and superiors,
and so on.'' Floyd's imaginary men were essentially carrying out coroutines,
but their actions could not be described naturally in any programming
languages that were available to Floyd when he wrote about the subject in
1964, so he presented the algorithm as a flow chart.
Ole-Johan Dahl later gave an elegant implementation of Floyd's algorithm
using the features of {\mc SIMULA~67}, in \S2.1.2 of [\Dsyn].

The coroutine concept was refined further during the 1970s; see, for example,
[\VM] and the references cited therein. But today's programming languages
have replaced those ideas with more modern notions such as ``threads'' and
``closures,'' which (while admirable in themselves) support coroutines only
in a rather awkward and cumbersome manner. The simple principles of old-style
coroutines, which Dahl called {\it quasi-parallel processes}, deserve to be
resurrected again and given better treatment by the programming languages of
tomorrow.

In this paper we will study examples for which a well-designed
compiler could transform certain families of coroutines
into optimized code, just as compilers
can often transform recursive procedures into iterative routines that require
less space and/or time.

The ideas presented below were motivated by applications to
the exhaustive generation of combinatorial objects. For example,
consider a coroutine that wants to look at all permutations
of $n$~elements; it can call repeatedly on a
permutation-generation coroutine to produce the successive
arrangements. The latter coroutine repeatedly forms a new
permutation and calls on the former coroutine to inspect the
result. The permutation coroutine has its own internal
state\dash---its own local variables and its current location in
an ongoing computational process\dash---so it does not consider
itself to be a ``subroutine'' of the inspection coroutine.
The permutation coroutine might also invoke other
coroutines, which in turn are computational objects with
their own internal states.

We shall consider the problem of generating all $n$-tuples
$a_1a_2\ldots a_n$ of~0s and~1s with the property that $a_j
\le a_k$ whenever $j\to k$ is an arc in a given directed
graph. Thus $a_j=1$ implies that $a_k$ must also be~1;
if $a_k=0$, so is~$a_j$.
These $n$-tuples are supposed to form a ``Gray path,''
in the sense that only one bit~$a_j$ should change at each
step. For example, if $n=3$ and if we require $a_1\le a_3$ and $a_2\le a_3$,
five binary strings $a_1a_2a_3$ satisfy the inequalities, and one such
Gray path is
$$000,\ 001,\ 011,\ 111,\ 101.$$

The general problem just stated does not always have a
solution. For example, suppose the given digraph is 
$$\vcenter
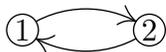
{\epsfbox{deco.1}}$$
so that the inequalities are $a_1\le a_2$ and $a_2\le a_1$;
then we are asking for a way to generate the tuples~00
and~11 by changing only one bit at a time, and this is
clearly impossible. Even if we stipulate that the digraph of
inequalities should contain no directed cycles,
we might encounter an example like
$$\vcenter
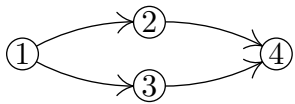
{\epsfbox{deco.2}}\,\,,$$
in which the Gray constraint cannot be achieved; here
the corresponding 4-tuples
$$0000, 0001, 0011, 0101, 0111, 1111$$
include four of even weight and two of odd weight, but a Gray
path must alternate between even and odd. Reasonably efficient methods
for solving the problem without Grayness are known
[\Squire, \Steiner], but we want to insist on single-bit
changes.

We will prove constructively that Gray paths always do exist if
we restrict consideration to directed graphs that are 
{\it totally acyclic}, in the sense that they contain no
cycles even if the directions of the arcs are ignored. Every
component of such a graph is a free tree in which a
direction has been assigned to each branch between two
vertices. Such digraphs are called {\it spiders}, because of
their resemblance to arachnids:
$$\epsfxsize=2in 
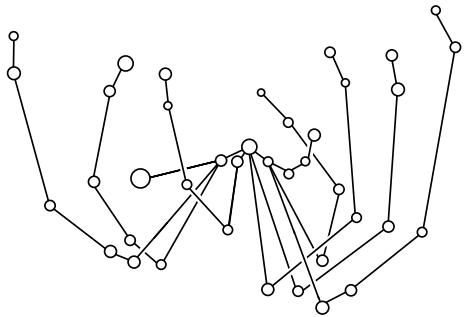
\vbox{\epsfbox{spider.eps}}$$
(In this diagram, as in others below, we assume that all
arcs are directed upwards. More complicated graph-theoretical spiders have
legs that change directions many more times than real spider legs do.)
The general problem of finding
all $a_1\ldots a_n$ such that 
$a_j \leq a_k$ when $j\to k$ in such a digraph is formally
called the task of ``generating the order ideals of an acyclic
poset''; it also is called, informally, ``spider squishing.''

Sections 1--3 of this paper discuss simple examples of the
problem in preparation for Section~4, which presents a
constructive proof that suitable Gray paths always exist.
The proof of Section~4 is implemented with coroutines in
Section~5, and Section~6 discusses the nontrivial task of
getting all the coroutines properly launched.

Section 7 describes a simple technique that is often able to
improve the running time. A generalization of that
technique leads in Section~8 to an efficient coroutine-free
implementation. Additional optimizations, which can be used
to construct an algorithm for the spider-squishing problem that is actually
{\it loopless}, are discussed in Section~9. (A loopless algorithm
needs only constant time to change each $n$-tuple to its successor.)

Section 10 concludes the paper and mentions several open problems
connected to related work.

\section 1. The unrestricted case. Let's begin by imagining
an array of friendly trolls called $T_1$, $T_2$, \dots,~$T_n$.
Each troll carries a lamp that is either off or on;
he also can be either awake or asleep.
Initially all the trolls are awake, and all their lamps are off.

Changes occur to the system when a troll is ``poked,''
according to the following simple rules: If $T_k$~is poked
\vadjust{\goodbreak}%
when he is awake, he changes the state of his lamp from off
to on or vice versa; then he becomes tired and goes to
sleep. Later, when the sleeping~$T_k$ is poked again, he
wakes up and pokes his left neighbor $T_{k-1}$, without making any change to
his own lamp. (The leftmost troll $T_1$ has no left neighbor, so he simply
awakens when poked.)

At periodic intervals an external driving force~$D$ pokes
the rightmost troll~$T_n$, initiating a chain of events that
culminates in one lamp changing its state. The process
begins as follows, if we use the digits~0 and~1 to represent
lamps that are respectively off or on, and if we underline
the digit of a sleeping troll:
$$\vbox{\halign{\dots#\quad&#\hfil\cr
               0000&Initial state\cr
             000\b1&$D$ pokes $T_n$\cr
             00\b11&$D$ pokes $T_n$, who wakes up and pokes
$T_{n-1}$\cr
           00\b1\b0&$D$ pokes $T_n$\cr
             0\b110&$D$ pokes $T_n$, who pokes $T_{n-1}$, who
pokes $T_{n-2}$\cr
           0\b11\b1&$D$ pokes $T_n$\cr
           0\b1\b01&$D$ pokes $T_n$, who pokes $T_{n-1}$\cr}}$$
The sequence of underlined versus not-underlined digits acts essentially
as a binary counter. And the sequence of digit patterns, in which
exactly one bit changes at each step, is a {\it Gray
binary\/} counter, which follows the well-known Gray binary
code; it also corresponds to the process of replacing rings
in the classic Chinese ring puzzle [\Kiv]. Therefore the
array of trolls solves our problem of generating all $n$-tuples
$a_1a_2\ldots a_n$, in the special case when the spider digraph has no arcs.
(This troll-oriented way to generate Gray binary code was
presented by the first author in a lecture at the University
of Oslo in October, 1972 [\KO].)

During the first $2^n$ steps of the process just described,
troll~$T_n$ is poked $2^n$ times,
troll $T_{n-1}$ is poked $2^{n-1}$ times, \dots, and troll~$T_1$ is
poked twice. The last step is special because $T_1$ has no left neighbor;
when he is poked the second time, all the trolls wake up, but no lamps
change. The driver~$D$ would like to know about this exceptional case, so we
will assume that $T_n$ sends a message to~$D$ after being poked, saying
`\\{true}' if one of the lamps has changed, otherwise saying `\\{false}'.
Similarly, if $1\le k<n$, $T_k$ will send a message to $T_{k+1}$ after being
poked, saying `\\{true}' if and only if one of
the first $k$ lamps has just changed state.

These hypothetical trolls $T_1$, \dots, $T_n$ correspond to $n$
almost-identical coroutines \\{poke}[1], \dots, $\\{poke}[n]$, whose actions
can be expressed in an ad hoc Algol-like language as follows:
\beginprogram
               Boolean coroutine $\\{poke}[k]$;\cr
              \quad while \\{true} do begin\cr
\qquad\label awake: $a[k]:=1-a[k]$; return $\\{true}$;\cr
\qquad\label asleep: if $k>1$ then return $\\{poke}[k-1]$
else return $\\{false}$;\cr
\qquad end\rm.\cr
\endprogram
\noindent Coroutine \\{poke}$[k]$ describes the action of~$T_k$,
implicitly retaining its own state of wakefulness: When \\{poke}$[k]$
is next activated after having executed the statement `{\bf return} \\{true}'
it will resume its program at label `asleep'; and it will resume at
label `awake' when it is next activated after `{\bf return} $\\{poke}[k-1]$'
or `{\bf return} \\{false}'.

In this example and in all the coroutine programs below, the enclosing
`{\bf while} \\{true} {\bf do begin} $\langle@P@\rangle$ {\bf end}' merely
says that program $\langle@P@\rangle$ should be repeated endlessly;
all coroutines that we shall encounter in this paper are immortal. (This is
fortunate, because Dahl [\Dsemi] has observed that proofs of correctness tend
to be much simpler in such cases.)

Our coroutines will also always be ``ultra-lightweight'' processes, in the
sense that they need no internal stack. They need only remember their current
positions within their respective programs, along with a few local variables
in some cases, together with the global ``lamp'' variables $a[1]$,
\dots,~$a[n]$. We can implement them using a single stack, essentially as if
we were implementing recursive procedures in the normal way,
pushing the address of a return point within~$A$ onto the stack when
coroutine~$A$ invokes coroutine~$B$, and resuming~$A$ after $B$ executes
a~{\bf return}. (Wang and Dahl~[\WD] used the term ``semicoroutine'' for this
special case. We are, however, using {\bf return} statements to return a
value, instead of using global variables for communication and saying `{\bf
detach}' as Wang and Dahl did.)
The only difference between our coroutine conventions
and ordinary subroutine actions is that a newly invoked coroutine always
begins at the point following its most recent {\bf return}, regardless of who
had previously invoked it. No coroutine will appear on the execution stack
more than once at any time.

Thus, for example, the coroutines \\{poke}[1] and \\{poke}[2] behave as
follows when $n=2$:
$$\vbox{\halign{#\quad&#\hfil\cr
     00&Initial state\cr
   0\b1&$\\{poke}[2]=\\{true}$\cr
   \b11&$\\{poke}[2]=\\{poke}[1]=\\{true}$\cr
 \b1\b0&$\\{poke}[2]=\\{true}$\cr
     10&$\\{poke}[2]=\\{poke}[1]=\\{false}$\cr
   1\b1&$\\{poke}[2]=\\{true}$\cr
   \b01&$\\{poke}[2]=\\{poke}[1]=\\{true}$\cr
 \b0\b0&$\\{poke}[2]=\\{true}$\cr
     00&$\\{poke}[2]=\\{poke}[1]=\\{false}$\cr}}$$
The same cycle will repeat indefinitely, because
everything has returned to its initial state.

Notice that the repeating cycle in this example consists of
two distinct parts. The first half cycle, before $\\{false}$
is returned, generates all two-bit patterns in Gray binary
order $(00,01,11,10)$; the other half generates those
patterns again, but in the {\it reverse\/} order $(10,11,01,00)$.
Such behavior will be characteristic of all the coroutines that we
shall consider for the spider-squishing problem: Their task
will be to run through all $n$-tuples $a_1\ldots a_n$ such
that $a_j \leq a_k$ for certain given pairs $(j,k)$, always
returning \\{true} until all permissible patterns have been generated;
then they are supposed to run through those $n$-tuples again
in reverse order, and to repeat the process ad infinitum. 

Under these conventions, a driver program of the following form
will cycle through the answers,
printing a line of dashes between each complete listing:
\beginprogram
\rm$\langle\,$Create all the coroutines$\,\rangle$;\cr
\rm$\langle\,$Put each lamp and each coroutine into the proper initial
 state$\,\rangle$;\cr
while $\\{true}$ do begin\cr
 \quad for $k:=1$ step 1 until $n$ do $\\{write}
(a[k])$;\cr
 \quad $\\{write}(\\{newline})$;\cr 
 \quad if not \\{root} then
$\\{write}(\hbox{\tt"-----"}, \\{newline})$;\cr
\quad end\rm.\cr
\endprogram
Here \\{root} denotes a coroutine that can potentially activate all the
others; for example, \\{root} is $\\{poke}[n]$ in the particular case
that we've been considering.
In practice, of course, the driver would normally carry out
some interesting process on the bits $a_1\ldots a_n$,
instead of merely outputting them to a file.

The fact that coroutines \\{poke}[1], \dots, $\\{poke}[n]$ do indeed generate
Gray binary code is easy to verify by induction on~$n$. The case $n=1$ is
trivial, because the outputs will clearly be
$$\vcenter{\halign{\tt#\cr
0\cr
1\cr
-----\cr
1\cr
0\cr
-----\cr}}$$
and so on. On the other hand
if $n>1$, assume that the successive contents of $a_1\ldots
a_{n-1}$ are $\alpha_0$, $\alpha_1$, $\alpha_2$, \dots\ when we repeatedly
invoke $\\{poke}[n-1]$, assuming that $\alpha_0=0\ldots0$ and that all
coroutines are initially at the label `awake'; assume further that \\{false}
is returned just before $\alpha_m$ when $m$ is a multiple of~$2^{n-1}$,
otherwise the returned value is~\\{true}. Then repeated invocations of
$\\{poke}[n]$ will lead to the successive lamp patterns
$$\alpha_00,\ \alpha_01,\ \alpha_11,\ \alpha_10,\ \alpha_20,\ \alpha_21,\
\ldots,$$
and \\{false} will be returned after every sequence of $2^n$ outputs.
These are precisely the
patterns of $n$-bit Gray binary code, alternately in forward order and
reverse order.

\section 2. Chains. Now let's go
to the opposite extreme and suppose that the digraph of
constraints is an oriented path or chain,
$$1\to2\to\cdots\to n@.$$
In other words, we want now to generate all $n$-tuples
$a_1a_2\ldots a_n$ such that
$$0\le a_1 \leq a_2\le\cdots \leq a_n \le1,$$
proceeding alternately forward and backward in Gray order.
Of course this problem is trivial, but we want to do it with
coroutines so that we'll be able to tackle more difficult
problems later.

Here are some coroutines that do the new job, if the driver program initiates
action by invoking the root coroutine \\{bump}[1]:
\beginprogram
Boolean coroutine $\\{bump}[k]$;\cr
\quad while $\\{true}$ do begin\cr
\qquad \label awake0: if $k<n$ then while $\\{bump}[k+1]$ do
return $\\{true}$;\cr
\qquad $a[k]:=1$; return $\\{true}$;\cr
\qquad \label asleep1: return $\\{false}$;\
 comment $a_k\ldots a_n=1\ldots1$;\cr
\qquad \label awake1: $a[k]:=0;$ return $\\{true}$;\cr
\qquad \label asleep0: if $k<n$ then while $\\{bump}[k+1]$ do
return $\\{true}$;\cr 
\qquad return $\\{false}$;\
 comment $a_k\ldots a_n=0\ldots0$;\cr
\qquad end\rm.\cr
\endprogram
For example, the process plays out as follows when $n=3$:
$$\vbox{\halign{#\quad&#\hfil\quad&#\hfil\cr
      000&Initial state & 123\cr
    00\b1&$\\{bump}[1]=\\{bump}[2]=\\{bump}[3]=\\{true}$&12\b3\cr
    0\b11&$\\{bump}[1]=\\{bump}[2]=\\{true}$, $\\{bump}[3]=\\{false}$&1\b2\cr
    \b111&$\\{bump}[1]=\\{true}$, $\\{bump}[2]=\\{false}$&\b1\cr
      111&$\\{bump}[1]=\\{false}$&1\cr
    \b011&$\\{bump}[1]=\\{true}$&\b12\cr
  \b0\b01&$\\{bump}[1]=\\{bump}[2]=\\{true}$&\b1\b23\cr
\b0\b0\b0&$\\{bump}[1]=\\{bump}[2]=\\{bump}[3]=\\{true}$&\b1\b2\b3\cr
      000&$\\{bump}[1]=\\{bump}[2]=\\{bump}[3]=\\{false}$&123\cr}}$$
Each troll's action now depends on whether his lamp is lit
as well as on his state of wakefulness. A troll with an
unlighted lamp always passes each bump to the right, without
taking any notice unless a $\\{false}$ reply comes back. In
the latter case, he acts as if his lamp had been
lit\dash---namely, he either returns $\\{false}$ (if just
awakened), or he changes the lamp, returns 
$\\{true}$, and nods off.
The Boolean value returned in each case is \\{true} if and only if a lamp has
changed its state during the current invocation of $\\{bump}[k]$.

({\it Note:\/} The numbers `123', `12\b3', \dots\ at the right
of this example correspond to an encoding that will be
explained in Section~8 below. A similar column of somewhat
inscrutable figures will be given with other examples we
will see later, so that the principles of Section~8 will be
easier to understand when we reach that part of the story.
There is no need to decipher such notations until then; all
will be revealed eventually.)

The dual situation, in which all inequalities are reversed so
that we generate all 
$a_1a_2\ldots a_n$ with 
$$1\ge a_1\ge a_2\ge\cdots\ge a_n\ge0,$$
can be implemented by interchanging the roles of 0 and 1 and starting
the previous sequence in the midpoint of its period:
\beginprogram
\quad Boolean coroutine $\\{cobump}[k]$;\cr
\qquad while $\\{true}$ do begin\cr
\qquad\quad\label awake0: $a[k]:=1$; return $\\{true}$;\cr
\qquad\quad\label asleep1: if $k<n$ then while $\\{cobump}
[k+1]$ do return $\\{true}$;\cr
\qquad\quad return $\\{false}$;\
 comment $a_k\ldots a_n=1\ldots1$;\cr
\qquad\quad\label awake1: if $k<n$ then while $\\{cobump}
[k+1]$ do return $\\{true}$;\cr
\qquad\quad $a[k]:=0@$; return $\\{true}$;\cr
\qquad\quad\label asleep0: return $\\{false}$;\
 comment $a_k\ldots a_n=0\ldots0$;\cr
\qquad\quad end\rm.\cr
\endprogram

A mixed situation in which the constraints are
$$0 \leq a_n \leq a_{n-1} \leq\cdots \leq a_{m+1} \leq a_1
\leq a_2 \leq\cdots \leq 
a_m \le1$$
is also worthy of note. Again the underlying digraph is a
chain, and the driver repeatedly bumps troll $T_1$; but when
$1<m<n$, the coroutines are a mixture of those we've just
seen:

\goodbreak
\noindent\beginwideprogram
\quad Boolean coroutine $\\{mbump}[k]$;\cr
\qquad while $\\{true}$ do begin\cr
\qquad\quad \label awake0: if $k < m$ then while
 $\\{mbump}[k+1]$ do return $\\{true}$;\cr
\qquad\quad $a[k]:=1$; return $\\{true}$;\cr
\qquad\quad \label asleep1: if $m<k\;\wedge\;k<n$
 then while $\\{mbump}[k{+}1])$ do return $\\{true}$;\cr
\qquad\quad if $k=1\;\wedge\;m<n$ then while $\\{mbump}[m{+}1])$ do
return $\\{true}$;\cr
\qquad\quad return $\\{false}$;\cr
\qquad\quad \label awake1: if $m<k\;\wedge\;k<n$
 then while $\\{mbump}[k{+}1])$ do return $\\{true}$;\cr
\qquad\quad if $k=1\;\wedge\;m<n$ then while $\\{mbump}[m{+}1])$ do
return $\\{true}$;\cr
\qquad\quad $a[k]:=0@$; return $\\{true}$;\cr
\qquad\quad \label asleep0: if $k < m$ then while
$\\{mbump}[k+1]$ do return $\\{true}$;\cr
\qquad\quad return $\\{false}$;\cr
\qquad\quad end\rm.\cr
\endprogram
The reader is encouraged to simulate the $\\{mbump}$ coroutines by hand
when, say, $m=2$ and $n=4$, in order to develop a better intuition about
coroutine behavior. Notice that when
$m\approx{1\over2}n$, signals need to propagate only about
half as far as they do when $m=1$ or $m=n$.

Still another simple but significant variant arises when
several separate chains are present. The digraph might, for
example, be
$$\advance\abovedisplayskip-.7\baselineskip
\advance\belowdisplayskip-.3\baselineskip
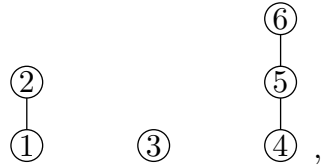
\hbox{\epsfbox{deco.3}\enspace\raise.5ex\hbox{,}}$$
in which case we want all $6$-tuples of bits $a_1\ldots a_6$
with $a_1 \leq a_2$ and 
$a_4 \leq a_5 \leq a_6$. In general, suppose there is a set
of endpoints $E=\{e_1,\ldots,e_m\}$ such that
$$1=e_1<\cdots<e_m \leq n,$$
and we want
$$a_k\in\{0,1\}\quad\hbox{for $1 \leq k \leq n$};\qquad
a_{k-1} \leq a_k\quad\hbox{for $k\notin E$}.$$
(The set $E$ is $\{1,3,4\}$ in the example shown.)
The following coroutines $\\{ebump}[k]$, for $1 \leq k \leq
n$, generate all such $n$-tuples if the driver invokes
$\\{ebump}[e_m]$:
\beginprogram
\quad Boolean coroutine $\\{ebump}[k]$;\cr
\qquad while $\\{true}$ do begin\cr
\qquad\label awake0: if $k+1\notin E\cup\{n+1\}$ then while
$\\{ebump}[k+1]$ do return $\\{true}$;\cr
\qquad $a[k]:=1;$ return $\\{true}$;\cr
\qquad\label asleep1: if $k\in E\setminus\{1\}$ then return
$\\{ebump}[k']$ else return 
$\\{false}$;\cr
\qquad\label awake1: $a[k]:=0;$ return $\\{true}$;\cr
\qquad\label asleep0: if $k+1\notin E\cup\{n+1\}$ then while
$\\{ebump}[k+1]$ do return $\\{true}$;\cr
\qquad if $k\in E\setminus\{1\}$ then return $\\{ebump}[k']$
else return $\\{false}$;\cr
\qquad end\rm.\cr
\endprogram
Here $k'$ stands for $e_{j-1}$ when $k=e_j$ and $j>1$.
These routines reduce to $\\{poke}$ when
$E=\{1,2,\ldots,n\}$ and to $\\{bump}$ when 
$E=\{1\}$.
If $E=\{1,3,4\}$, they will generate all 24 bit patterns
such that $a_1 \leq a_2$ and $a_4 \leq a_5 \leq a_6$ in the
order
$$\openup1\jot
\vbox{\halign{\hfil#\hfil\ &\hfil#\hfil\ &\hfil#\hfil\
&\hfil#\hfil\ &\hfil#\hfil\ &\hfil#\hfil\ &\hfil#\hfil\
&\hfil#\hfil\cr
000000,&00000\b1,&0000\b11,&000\b111,&00\b1111,&00\b1\b011,&00\b1\b0\b01,&00\b1\b0\b0\b0,\cr
0\b11000,&0\b1100\b1,&0\b110\b11,&0\b11\b111,&0\b1\b0111,&0\b1\b0\b011,&0\b1\b0\b0\b01,&0\b1\b0\b0\b0\b0,\cr
\b110000,&\b11000\b1,&\b1100\b11,&\b110\b111,&\b11\b1111,&\b11\b1\b011,&\b11\b1\b0\b01,&\b11\b1\b0\b0\b0;\cr}}$$
then the sequence will reverse itself:
$$\openup1\jot
\vbox{\halign{\hfil#\hfil\ &\hfil#\hfil\ &\hfil#\hfil\
&\hfil#\hfil\ &\hfil#\hfil\ &\hfil#\hfil\ &\hfil#\hfil\
&\hfil#\hfil\cr
111000,&11100\b1,&1110\b11,&111\b111,&11\b0111,&11\b0\b011,&11\b0\b0\b01,&11\b0\b0\b0\b0,\cr
\b010000,&\b01000\b1,&\b0100\b11,&\b010\b111,&\b01\b1111,&\b01\b1\b011,&\b01\b1\b0\b01&\b01\b1\b0\b0\b0,\cr
\b0\b01000,&\b0\b0100\b1,&\b0\b010\b11,&\b0\b01\b111,&\b0\b0\b0111,&\b0\b0\b0\b011,&
\b0\b0\b0\b0\b01,&\b0\b0\b0\b0\b0\b0.\cr}}$$

In our examples so far we have discussed several families of
cooperating coroutines and claimed that they generate
certain $n$-tuples, but we haven't proved anything
rigorously.  A formal theory of coroutine semantics is
beyond the scope of this paper, but we should at least try
to construct a semi-formal demonstration that $\\{ebump}$ is
correct.

The proof is by induction on $|E|$, the number of chains.
If $|E|=1$, $\\{ebump}[k]$ reduces to $\\{bump}[k]$, and
we can argue by induction on $n$.  The result is obvious
when $n=1$.  If $n>1$, suppose repeated calls on $\\{bump}
[2]$ cause $a_2\ldots a_n$ to run through the $(n-1)$-tuples
$\alpha_0$, $\alpha_1$, $\alpha_2$, \dots, where $\\{bump}[2]$
is $\\{false}$ when it produces $\alpha_t=\alpha_{t-1}$.
Such a repetition will occur if and only if $t$~is a
multiple of~$n$, because $n$ is the number of distinct $(n-1)$-tuples with
$a_2 \leq\cdots \leq a_n$.  We know by induction that the
sequence has reflective symmetry:
$\alpha_j=\alpha_{2n-1-j}$ for $0 \leq j < n$.
Furthermore, $\alpha_{j+2n}=\alpha_j$ for all $j\ge0$.  To
complete the proof we observe that repeated calls on
$\\{bump}[1]$ will produce the 
$n$-tuples
$$\eqalign{&0\alpha_0,\     0\alpha_1,\                  \ldots,\ 
0\alpha_{n-1},\  \b1\alpha_n,\cr
           &1\alpha_n,\   \b0\alpha_n,\  \b0\alpha_{n+1},\ \ldots,\ 
\b0\alpha_{2n-1},\cr
           &0\alpha_{2n},\  0\alpha_{2n+1},\             \ldots,\ 
0\alpha_{3n-1},\ \b1\alpha_{3n},\cr}$$
and so on, returning $\\{false}$ every $(n+1)^{\rm st}$ step
as desired.

If $|E|>1$, let $E=\{e_1,\ldots, e_m\}$, so that
$e'_m=e_{m-1}$, and suppose that repeated calls on
$\\{ebump}[e_{m-1}]$ produce the $(e_m-1)$-tuples
$\alpha_0$, $\alpha_1$, $\alpha_2$, \dots$@$.  Also suppose that
calls on $\\{ebump}[e_m]$ would set the remaining bits
$a_{e_m}\ldots a_n$ to the $(n+1-e_m)$-tuples
$\beta_0$, $\beta_1$, $\beta_2$, \dots, if $E$ were
empty instead of $\{e_1,\ldots,e_m\}$;
this sequence $\beta_0$, $\beta_1$, $\beta_2$, \dots\ is like the
output of $\\{bump}$.  The $\alpha$ and $\beta$ sequences
are periodic, with respective periods of length $2M$ and $2N$ for some
$M$ and $N$; they also have reflective symmetry
$\alpha_j=\alpha_{@2M-1-j}$, $\beta_k=\beta_{@2N-1-k}$.  It
follows that 
$\\{ebump}[e_m]$ is correct, because it produces the
sequence
$$\eqalign{\gamma_0,\gamma_1,\gamma_2,\ldots={}&\alpha_0\beta_0,\ 
\alpha_0\beta_1,\ \ldots,\ \alpha_0\beta_{N-1},\cr
&\alpha_1\beta_N,\ \alpha_1\beta_{N+1},\ \ldots,\ \alpha_1\beta_{@2N-1},\cr
&\quad\vdots\cr
&\alpha_{M-1}\beta_{(M-1)N},\ \alpha_{M-1}\beta_{(M-1)N+1},\ \ldots,\  
\alpha_{M-1}\beta_{MN-1},\cr
&\alpha_M\beta_{MN},\ \alpha_M\beta_{MN+1},\ \ldots,\ 
\alpha_M\beta_{(M+1)N-1},\cr
&\quad\vdots\cr
&\alpha_{@2M-1}\beta_{(2M-1)N},\ \alpha_{@2M-1}\beta_{(2M-1)N+1},\ 
\ldots,\ \alpha_{@2M-1}\beta_{@2MN-1},\ \ldots\cr}$$
which has period length $2M\mskip-2muN$ and satisfies
$$\gamma_{Nj+k}=\alpha_j\beta_{Nj+k}=\alpha_{@2M-1-j} 
\beta_{@2MN-1-Nj-k}=\gamma_{@2MN-1-Nj-k}$$
for $0 \leq j<M$ and $0 \leq k<N$.

The patterns output by $\\{ebump}$ are therefore easily seen
to be essentially the same as the so-called {\it reflected
Gray paths\/} for radices 
$e_2+1-e_1$, \dots, $e_m+1-e_{m-1}$, $n+2-e_m$ (see [\Kiv]); the
total number of outputs is 
$$(e_2+1-e_1)\ldots(e_m+1-e_{m-1})(n+2-e_m).$$

\section 3. Ups and downs.  Now let's consider a ``fence''
digraph
$$\vcenter
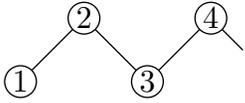
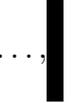
{\epsfbox{deco.4}}\enspace\ldots,$$
which leads to $n$-tuples that satisfy the up-down constraints
$$a_1 \leq a_2\ge a_3 \leq a_4\ge\cdots.$$
A reasonably simple set of coroutines can be shown to handle
this case, rooted at $\\{nudge}[1]$:
\beginprogram
\quad Boolean coroutine $\\{nudge}[k]$;\cr
\qquad while $\\{true}$ do begin\cr
\qquad\quad\label awake0: if $k' \leq n$ then while $\\{nudge}
[k']$ do return $\\{true}$;\cr
\qquad\quad $a[k]:=1$; return $\\{true}$;\cr
\qquad\quad\label asleep1: if $k'' \leq n$ then while $\\{nudge}
[k'']$ do return $\\{true}$;\cr
\qquad\quad return $\\{false}$;\cr
\qquad\quad\label awake1: if $k'' \leq n$ then while $\\{nudge}[k'']$
do return $\\{true}$;\cr
\qquad\quad $a[k]:=0@$; return $\\{true}$;\cr
\qquad\quad\label asleep0: if $k' \leq n$ then while $\\{nudge}
[k']$ do return $\\{true}$;\cr
\qquad\quad return $\\{false}$;\cr
\qquad\quad end\rm.\cr
\endprogram
Here $(k',k'')=(k+1,k+2)$ when $k$ is odd, $(k+2,k+1)$ when $k$
is even. But these 
coroutines do {\it not\/} work when they all begin at `awake0'
with $a_1a_2\ldots a_n=00\ldots0@$; 
they need to be initialized carefully. For example, when
$n=6$ it turns out that exactly eleven
patterns of odd weight need to be generated, and exactly ten
patterns of even weight, so a Gray 
path cannot begin or end with an even-weight pattern such as
000000 or 111111. One
proper starting configuration is obtained if we set $a_1\ldots 
a_n$ to the first $n$ bits of the infinite string $000111000111\ldots{}$,
and if we start coroutine $\\{nudge}[k]$ 
at `awake0' if $a_k=0$, at `awake1' if $a_k=1$.
For example, the sequence of results 
when $n=4$ is 
$$\vbox{\halign{#&\quad#\hfil\quad&#\cr
0001 &Initial configuration &124\cr
000\b0 &$\\{nudge}[1]=\\{nudge}[2]=\\{nudge}[4]=\\{true}$
&12\b4\cr
0\b100 &$\\{nudge}[1]=\\{nudge}[2]=\\{true}$, $\\{nudge}
[4]=\\{false}$ &1\b234\cr
0\b10\b1&$\\{nudge}[1]=\\{nudge}[2]=\\{nudge}
[3]=\\{nudge}[4]=\\{true}$ &1\b23\b4\cr
0\b1\b11&$\\{nudge}[1]=\\{nudge}[2]=\\{nudge}
[3]=\\{true}$, $\\{nudge}[4]=\\{false}$ 
&1\b2\b3\cr
\b1111&$\\{nudge}[1]=\\{true}$, $\\{nudge}
[2]=\\{nudge}[3]=\\{false}$ &\b13\cr
\b11\b01&$\\{nudge}[1]=\\{nudge}[3]=\\{true}$ &\b1\b34\cr
\b11\b0\b0&$\\{nudge}[1]=\\{nudge}[3]=\\{nudge}
[4]=\\{true}$ &\b1\b3\b4\cr
1100&$\\{nudge}[1]=\\{nudge}[3]=\\{nudge}[4]=\\{false}$
&134\cr
110\b1&$\\{nudge}[1]=\\{nudge}[3]=\\{nudge}[4]=\\{true}$&13\b4\cr
11\b11&$\\{nudge}[1]=\\{nudge}[3]=\\{true}$,
$\\{nudge}[4]=\\{false}$&1\b3\cr
\b0111&$\\{nudge}[1]=\\{true}$,
$\\{nudge}[3]=\\{false}$&\b123\cr
\b01\b01&$\\{nudge}[1]=\\{nudge}[2]=\\{nudge}
[3]=\\{true}$&\b12\b34\cr
\b01\b0\b0&$\\{nudge}[1]=\\{nudge}[2]=\\{nudge}
[3]=\\{nudge}[4]=\\{true}$ 
&\b12\b3\b4\cr
\b0\b000&$\\{nudge}[1]=\\{nudge}[2]=\\{true}$, $\\{nudge}
[3]=\\{nudge}[4]=\\{false}$ 
&\b1\b24\cr
\b0\b00\b1&$\\{nudge}[1]=\\{nudge}[2]=\\{nudge}
[4]=\\{true}$&\b1\b2\b4\cr
0001&$\\{nudge}[1]=\\{nudge}[2]=\\{nudge}
[4]=\\{false}$&124\cr}}$$
Again the cycle repeats with reflective symmetry; and again,
some cryptic notations appear that 
will be explained in Section~8.  The correctness of
$\\{nudge}$ will follow from results we shall 
prove later.

\section 4. The general case. We have seen that cleverly constructed
coroutines are able to generate Gray paths for several rather different
special cases of the spider-squishing problem; thus it is natural to hope
that similar techniques will work in the general case when
an arbitrary totally acyclic digraph is given.  The spider
$$
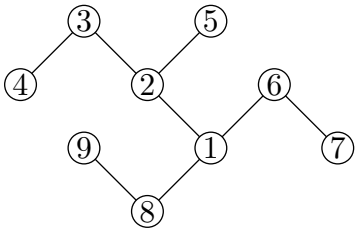
\vbox{\epsfbox{deco.5}}$$
illustrates most of the complications that might face us, so
we shall use it as a running example.  
In general we shall assume that the vertices have been
numbered in {\it preorder}, as defined in [\Ki, Section 2.3.2],
when the 
digraph is considered to be a forest (ignoring the arc
directions).  This means that the smallest vertex in each component is the
root of that component, and that all 
vertex numbers of a component are consecutive.  Furthermore,
the children of each node are 
immediately followed in the ordering by their descendants.
The descendants of each node~$k$ 
form a subspider consisting of nodes~$k$ through $\scope
(k)$, inclusive; we shall call this 
``spider $k$.''  For example, spider~2 consists of nodes
$\{2,3,4,5\}$, and $\scope (2)=5$.  Our 
sample spider has indeed been numbered in preorder, because it can
be drawn as a properly numbered 
tree with directed branches:
$$
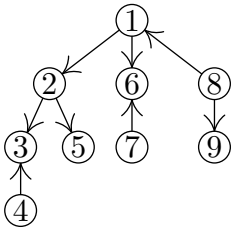
\vbox{\epsfbox{deco.6}}$$
The same spider could also have been numbered in many other ways, because any
vertex of the digraph could have been 
chosen to be the root, and because the resulting trees can
be embedded several ways into the 
plane by permuting the children of each family.

Assume for the moment that the digraph is connected; thus it
is a tree with root 1.  A nonroot 
vertex~$x$ is called {\it positive\/} if the path from~1
to~$x$ ends with an arc directed 
towards~$x$, {\it negative\/} if that path ends with an arc
directed away from~$x$.  Thus the 
example spider has positive vertices $\{2,3,5,6,9\}$ and
negative vertices $\{4,7,8\}$.

Let us write $x\to ^*y$ if there is a directed path from~$x$
to~$y$ in the digraph.  Removing all 
vertices~$x$ such that $x\to ^*1$ disconnects the graph into
a number of pieces having positive 
roots; in our example, the removal of $\{1,8\}$ leaves three
components rooted at $\{2,6,9\}$.  
We call these roots the {\it positive vertices near\/}~1, and
we denote that set by~$U_1$.  
Similarly, the {\it negative vertices near\/}~1 are obtained
when we remove all vertices~$y$ such 
that $1\to ^*y$; the set of resulting roots, denoted
by~$V_1$, is $\{4,7,8\}$ in our example, 
because we remove $\{1,2,3,5,6\}$.

The relevant bit patterns $a_1\ldots a_n$ for which $a_1=0$
are precisely those that we obtain if 
we set $a_j=0$ whenever $j\to ^*1$ and if we supply bit
patterns for each subspider rooted at 
a vertex of~$U_1$.  Similarly, the bit patterns for which
$a_1=1$ are precisely those we obtain 
by setting $a_k=1$ whenever $1\to ^*k$ and by supplying
patterns for each subspider rooted at a 
vertex of~$V_1$.
Thus if $n_k$ denotes the number of bit patterns for
spider~$k$, the total number of suitable 
patterns $a_1\ldots a_n$ is $\prod_{u\in U_1} n_u
+\prod_{v\in V_1} n_v$.

The sets $U_k$ and $V_k$ of positive and negative vertices
near~$k$ are defined in the same 
way for each spider~$k$.

Every positive child of~$k$ appears in~$U_k$, and every
negative child appears in~$V_k$.  
These are called the {\it principal\/} elements of~$U_k$
and~$V_k$.  Every nonprincipal member 
of~$U_k$ is a member of~$U_v$ for some unique principal
vertex~$v$ of~$V_k$.  Similarly, 
every nonprincipal member of~$V_k$ is a member of~$V_u$ for
some unique principal 
vertex~$u$ of~$U_k$.  For example, the principal members
of~$U_1$ are 2 and 6; the other 
member, 9, belongs to~$U_8$, where 8 is a principal member
of~$V_1$.

We will prove that the bit patterns $a_1\ldots a_n$ can
always be arranged in a Gray path such 
that bit~$a_1$ begins at 0 and ends at 1, changing exactly
once.  By induction, such paths exist 
for the~$n_u$ patterns in each spider~$u$ for $u\in U_1$.
And we can combine such paths 
into a single path that passes through all of the $\prod_{u\in U_1} n_u$
ways to combine those patterns, using a
reflected Gray code analogous to 
the output of $\\{ebump}$ in Section~3 above.  Thus, if we
set $a_k=0$ for all~$k$ such that 
$k\to ^*1$, we get a Gray path $P_1$ for all suitable
patterns with $a_1=0$.  Similarly we can 
construct a Gray path~$Q_1$ for the $\prod_{v\in V_1} n_v$
suitable patterns with $a_1=1$.  Thus, all
we need to do is prove that it is possible to
construct~$P_1$ and~$Q_1$ in such a way that the 
last pattern in~$P_1$ differs from the first pattern
of~$Q_1$ only in bit~$a_1$.  Then 
$G_1=(P_1,Q_1)$ will be a suitable Gray path that solves our problem.

For example, consider the subspiders for $U_1=\{2,6,9\}$ in
the example spider.  An inductive 
construction shows that they have respectively
$(n_2,n_6,n_9)=(8,3,2)$ patterns, with 
corresponding Gray paths
$$\eqalign{G_2&=0000, 0001, 0101, 0100, 0110, 0111, 1111,
1101;\cr
                   G_6&=00, 10, 11;\cr
                   G_9&=0,1.\cr}$$
We obtain 48 patterns~$P_1$ by setting $a_1=a_8=0$ and
using~$G_2$ for $a_2a_3a_4a_5$, 
$G_6$~for $a_6a_7$, and~$G_9$ for~$a_9$, taking care to end
with $a_2=a_6=1$.  Similarly, 
the subspiders for $V_1=\{4,7,8\}$ have
$(n_4,n_7,n_8)=(2,2,3)$ patterns, and paths 
$$\eqalign{G_4&=0,1;\cr
           G_7&=0,1;\cr
           G_8&=00,01,11.}$$
We obtain 12 patterns $Q_1$ by setting
$a_1=a_2=a_3=a_5=a_6=1$ and using~$G_4$ for~$a_4$, $G_7$
for~$a_7$, and $G_8$ for $a_8a_9$, taking care to begin with
$a_8=0$.  Combining these observations, we see that
$P_1$~should end with $011011100$, and $Q_1$~should begin
with $111011100$.

In general, the last element of $P_k$ and the first element
of~$Q_k$ can be determined as follows:  For all children~$j$
of~$k$, set $a_j\ldots a_{@\scope(j)}$ to the last element of
the previously computed Gray path~$G_j$ if $j$~is positive,
or to the first element of~$G_j$ if $j$~is negative.  Then
set $a_k=0$ in~$P_k$, $a_k=1$ in~$Q_k$.  It is easy
to verify that these rules make $a_j=0$ whenever $j\to^*k$,
and $a_j=1$ whenever $k\to^*j$,
for all~$j$ such that $k<j\le\scope(k)$.  A reflected Gray
code based on the paths~$G_u$ for $u\in U_k$ can be used to
construct~$P_k$ ending at the transition values, having
$a_k=0$; and $Q_k$~can be constructed from those starting
values based on the paths~$G_v$ for $v\in V_k$, having
$a_k=1$.  Thus we obtain a Gray path $G_k=(P_k,Q_k)$.

We have therefore constructed a Gray path for spider 1,
proving that the spider-squishing problem has a solution
when the underlying digraph is connected.  To complete the
construction for the general case, we can artificially
ensure that the graph is connected by introducing a new
vertex~0, with arcs from 0 to the roots of the components.
Then $P_0$~will be the desired Gray path, if we suppress
bit~$a_0$ (which is zero throughout~$P_0$).

\section 5.  Implementation via coroutines.  By constructing
families of sets~$U_k$ and~$V_k$ and identifying principal
vertices in those sets, we have shown the existence of a
Gray path for any given spider-squishing problem. Now let's
make the proof explicit by constructing a family of
coroutines that will generate the successive patterns
$a_1\ldots a_n$ dynamically, as in the examples worked out in
Sections~1--3 above.

\begingroup\advance\abovedisplayskip -.5\baselineskip 
           \advance\belowdisplayskip -.5\baselineskip 
First let's consider a basic substitution or ``plug-in'' operation
that applies to coroutines of the type we are using. Consider the
following coroutines $X$ and~$Y$:
\beginprogram
\quad Boolean coroutine $X$;\cr
\qquad while $\\{true}$ do begin\cr
\qquad\quad while $A$ do return $\\{true}$;\cr
\qquad\quad return $\\{false}$;\cr
\qquad\quad while $B$ do return $\\{false}$;\cr
\qquad\quad if $C$ then return $\\{true}$;\cr
\qquad\quad end;\cr
\noalign{\smallskip}
\quad Boolean coroutine $Y$;\cr
\qquad while $\\{true}$ do begin\cr
\qquad\quad while $X$ do return $\\{true}$;\cr
\qquad\quad return $Z$;\cr
\qquad\quad end\rm.\cr
\endprogram
Here $X$ is a more-or-less random coroutine that invokes three coroutines
$A$, $B$, $C$; coroutine $Y$ has a special structure that invokes $X$ and
an arbitrary coroutine $Z\ne X,Y$.  Clearly $Y$~carries out
essentially the same actions as the slightly faster
coroutine~\\{XZ} that we get from~$X$ by substituting $Z$
wherever $X$~returns $\\{false}$:
\beginprogram
\quad Boolean coroutine $\\{XZ}$;\cr
\qquad while $\\{true}$ do begin\cr
\qquad\quad while $A$ do return $\\{true}$;\cr
\qquad\quad return $Z$;\cr
\qquad\quad while $B$ do return $Z$;\cr
\qquad\quad if $C$ then return $\\{true}$;\cr
\qquad\quad end\rm.\cr
\endprogram

This plug-in principle applies in the same way whenever all {\bf
return} statements of $X$ are either 
`{\bf return} $\\{true}$' or `{\bf return} $\\{false}$'.
And we could cast $XZ$ into this same mold, if desired, by
writing `{\bf if $Z$ then return $\\{true}$ else return
$\\{false}$}' in place of `{\bf return} $Z$'.
\par\endgroup

In general we want to work with coroutines whose actions produce infinite
sequences $\alpha_1,\alpha_2,\ldots$ of period length~$2M$, where
$(\alpha_M,\ldots,\alpha_{@2M-1})$ is the reverse of
$(\alpha_0,\ldots,\alpha_{M-1})$, and where the coroutine
returns $\\{false}$ after producing~$\alpha_t$ if and only
if $t$~is a multiple of~$M$.  The proof at the end of
Section~2 shows that a construction like coroutine~$Y$ above, namely
\beginprogram
\quad Boolean coroutine $\\{AtimesB}$;\cr
\qquad while $\\{true}$ do begin\cr
\qquad\quad while $B$ do return $\\{true}$;\cr
\qquad\quad return $A$;\cr
\qquad\quad end\cr
\endprogram
yields a coroutine that produces such sequences of
period length~$2MN$ from coroutines~$A$ and~$B$ of
period lengths~$2M$ and~$2N$, when $A$ and $B$ affect disjoint bit positions
of the output sequences.

The following somewhat analogous coroutine produces such
sequences of period length $2(M+N)$:
\beginprogram
\quad Boolean coroutine $\\{AplusB}$;\cr
\qquad while $\\{true}$ do begin\cr
\qquad\quad while $A$ do return $\\{true}$;\cr
\qquad\quad $a[1]:=1$; return $\\{true}$;\cr
\qquad\quad while $B$ do return $\\{true}$;\cr
\qquad\quad return $\\{false}$;\cr
\qquad\quad while $B$ do return $\\{true}$;\cr
\qquad\quad $a[1]:=0@$; return $\\{true}$;\cr
\qquad\quad while $A$ do return $\\{true}$;\cr
\qquad\quad return $\\{false}$;\cr
\qquad\quad end\rm.\cr
\endprogram
This construction assumes that $A$ and $B$ individually generate
reflective periodic sequences $\alpha$ and~$\beta$ on bits
$a_2\ldots a_n$, and that $\alpha_M=\beta_0$.  The first half of
\\{AplusB} produces
$$0\alpha_0,\ \ldots,\ 0\alpha_{M-1},\ 1\beta_0,\ \ldots,\ 1\beta_{N-1},$$
and returns $\\{false}$ after forming~$1\beta_N$ (which equals $1\beta_{N-1}$).
The second half produces the $n$-tuples
$$1\beta_N,\ \ldots,\ 1\beta_{@2N-1},\ 0\alpha_M,\ \ldots,\  0\alpha_{@2M-1},$$
which are the first $M+N$ outputs in reverse; then it returns $\\{false}$,
after forming $0\alpha_{@2M}$ (which equals $0\alpha_0$).

The coroutines that we need to implement spider squishing
can be built up from variants of the primitive constructions
for product and sum just mentioned.  Consider the following
coroutines $gen [1]$, \dots, $gen[n]$, each of which receives
an integer parameter~$l$ whenever being invoked:
\beginprogram
\quad Boolean coroutine $\\{gen}[k](l)$; integer
$l$;\cr
\qquad while $\\{true}$ do begin\cr
\qquad\quad\label awake0: if $\\{maxu}[k]\ne0$ then while
$\\{gen}[\\{maxu}[k]](k)$ do return $\\{true}$;\cr
\qquad\quad $a[k]:=1$; return $\\{true}$;\cr
\qquad\quad\label asleep1: if $\\{maxv}[k]\ne0$ then while
$\\{gen}[\\{maxv}[k]](k)$ do return $\\{true}$;\cr
\qquad\quad if $\\{prev}[k]>l$ then return
$\\{gen}[\\{prev}[k]](l)$ else return 
$\\{false}$;\cr
\qquad\quad\label awake1: if $\\{maxv}[k]\ne0$ then while
$\\{gen}[\\{maxv}[k]](k)$ do return $\\{true}$;\cr
\qquad\quad $a[k]:=0@$; return $\\{true}$;\cr
\qquad\quad\label asleep0: if $\\{maxu}[k]\ne0$ then while
$\\{gen}[\\{maxu}[k]](k)$ do return $\\{true}$;\cr
\qquad\quad if $\\{prev}[k]>l$ then return
$\\{gen}[\\{prev}[k]](l)$ else return 
$\\{false}$;\cr
\qquad\quad end\rm.\cr
\endprogram
Here $\\{maxu}[k]$ denotes the largest element of~$U_k\cup\{0\}$, and
$\\{prev}[k]$ is a function that we shall define momentarily.
This function, like the sets~$U_k$ and~$V_k$, is statically
determined from the given totally acyclic digraph.

The idea of `$\\{prev}$' is that all elements of~$U_{@l}$ can
be listed as $u$, $\\{prev}[u]$, $\\{prev}\bigl[\\{prev}[u]\bigr]$, \dots,
until reaching an element $\leq l$, if we start with 
$u=\\{maxu}[l]$.  Similarly, all elements of~$V_{@l}$
can be listed as~$v$, $\\{prev}[v]$, $\\{prev}\bigl[\\{prev}[v]\bigr]$, \dots,
while those elements exceed~$l$, starting with 
$v=\\{maxv}[l]$.  The basic meaning of $\\{gen}[k]$ with
parameter~$l$ is to run through all bit patterns for the
spiders $u\le k$ in $U_{@l}$, if $k$~is a positive vertex, or for the
spiders $v\le k$ in~ $V_{@l}$, if vertex $k$~is negative.

The example spider of Section 4 will help clarify the
situation.  The following table shows the sets~$U_k$, $V_k$,
and a suitable function $\\{prev}[k]$, together with some
auxiliary functions by which $\\{prev}[k]$ can be determined in
general:
$$\vbox{\halign{\indent#\quad\hfil&\quad\hfil#\hfil&\quad\hfil#\hfil&\quad\hfil#\hfil&\quad\hfil#\hfil&\quad\hfil#\hfil&\quad\hfil#\hfil\cr
$k$&$\scope(k)$&   $U_k$   &   $V_k$
&$\\{prev}[k]$&ppro$(k)$&npro$(k)$\cr
 1 &    9      &$\{2,6,9\}$&$\{4,7,8\}$&     0    &   1
&   0   \cr
 2 &    5      & $\{3,5\}$ &  $\{4\}$  &     0    &   2
&   0   \cr
 3 &    4      &$\emptyset$&  $\{4\}$  &     0    &   3
&   0   \cr
 4 &    4      &$\emptyset$&$\emptyset$&     0    &   3
&   4   \cr
 5 &    5      &$\emptyset$&$\emptyset$&     3    &   5
&   0   \cr
 6 &    7      &$\emptyset$&  $\{7\}$  &     2    &   6
&   0   \cr
 7 &    7      &$\emptyset$&$\emptyset$&     4    &   6
&   7   \cr
 8 &    9      &  $\{9\}$  &$\emptyset$&     7    &   1
&   8   \cr
 9 &    9      &$\emptyset$&$\emptyset$&     6    &   9
&   8   \cr}}$$

If $u$ is a positive vertex, not a root, let $v_1$ be the
parent of~$u$.  Then if $v_1$~is negative, let~$v_2$ be the
parent of~$v_1$, and continue in this manner until reaching
a positive vertex~$v_t$, the nearest positive ancestor
of~$v_1$.  We call~$v_t$ the 
{\it positive progenitor\/} of~$v_1$, denoted ppro$(v_1)$. The
main point of this construction is that $u\in U_k$ if and
only if $k$~is one of the vertices 
$\{v_1,v_2,\ldots,v_t\}$.  Consequently
$$U_k\;=\;U_{@l}\cap\{k,k+1,\ldots,\scope(k)\}$$
if $l$~is the positive progenitor of~$k$.  Furthermore
$U_k$ and $U_{k'}$ are disjoint whenever $k$ and $k'$ are
distinct positive vertices.  Therefore we can define
$\\{prev}[u]$ for all positive nonroots~$u$ as the largest
element less than~$u$ in the set 
$U_k\cup\{0\}$, where $k={\rm ppro(parent}(u))$
is the positive progenitor of $u$'s parent.

Every element also has a negative progenitor, if we regard
the dummy vertex 0 as a negative vertex that is parent to
all the roots of the digraph.  Thus we define $\\{prev}[v]$ for
all negative~$v$ as the largest element less than~$v$ in the
set~$V_k\cup\{0\}$, where $k={\rm npro(parent}(v))$.

Notice that 9 is an element of both $U_1$ and~$U_8$ in the
example spider, so both 
$\\{gen}[9](1)$ and $\\{gen}[9](8)$ will be invoked at
various times.  The former will invoke $\\{gen}[6](1)$,
which will invoke $\\{gen}[2](1)$; the latter, however,
will merely flip bit~$a_9$ on and off, because $\\{prev}[9]$ does
not exceed~8.  There is only one coroutine $\\{gen}[9]$; its
parameter~$l$ is reassigned each time $\\{gen}[9]$ is
invoked.  (The two usages do not conflict, because
\vadjust{\goodbreak}
$\\{gen}[9](1)$ is invoked only when $a_1=0$, in which case
$a_8=0$ and $\\{gen}[8]$ cannot be active.)  Similarly,
$\\{gen}[4]$ can be invoked with $l=1,2$, or $3$; but in
this case there is no difference in behavior because 
$\\{prev}[4]=0$.

In order to see why $\\{gen}[k]$ works, let's consider first
what would happen if its parameter~$l$ were~$\infty$, so
that the test `$\\{prev}[k]>l@$' would always be false.  In
such a case $\\{gen}[k]$ is simply the \\{AplusB}
construction applied to 
$A=\\{gen}[\\{maxu}[k]](k)$ and $B=\\{gen}[\\{maxv}[k]](k)$.

On the other hand when 
$l$~is set to a number such that $k\in U_{@l}$ or $k\in
V_{@l}$, the coroutine $\\{gen}[k]$ is essentially the \\{AtimesB}
construction, because it results when $Z=\\{gen}[\\{prev}[k]](l)$ is
plugged in to the instance of \\{AplusB} that we've just discussed.
The effect is to obtain the Cartesian product of the sequence
generated with $l=\infty$ and the sequence generated by
$\\{gen}[\\{prev}[k]](l)$.

Thus we see that `{\bf if $\\{maxu}[k]\ne0$ then while
$\\{gen}[\\{maxu}[k]](k)$ do return $\\{true}$}'
generates the sequence~$P_k$ described in
Section~4, and `{\bf if $\\{maxv}\ne0$ then while $\\{gen}[\\{maxv}[k]]
(k)$ do return $\\{true}$}' generates~$Q_k$.  It follows
that $\\{gen}[k](\infty)$ generates the Gray path~$G_k$.
And we get the overall solution to our problem, path~$P_0$,
by invoking the root coroutine $\\{gen}[\\{maxu}[0]](0)$.

Well, there is one hitch:  Every time the \\{AplusB}
construction is used, we must be sure that coroutines~$A$
and~$B$ have been set up so that the last pattern of 
$A$ equals the first pattern of $B$.  We shall deal
with that problem in Section~6.

In the unconstrained case, when the given digraph has no arcs
whatsoever, we have 
$U_0=\{1,\ldots,n\}$ and all other $U$'s and $V$'s are
empty.  Thus $\\{prev}[k]=k-1$ for 
$1\le k\le n$, and $\\{gen}[k](0)$ reduces to the coroutine
$\\{poke}[k]$ of Section~1.

If the given digraph is the chain $1\to2\to\cdots\to n$, the
nonempty $U$'s and $V$'s are $U_k=\{k+1\}$ for $0\le k<n$.
Thus $\\{prev}[k]=0$ for all $k$, and $\\{gen}[k](l)$
reduces to the coroutine $\\{bump}[k]$ of Section~2.
Similar remarks apply to \\{cobump}, \\{mbump}, and \\{ebump}.

If the given digraph is the fence
$1\to2\gets3\to4\gets\cdots,$ we have $U_k=\{k'\}$ and
$V_k=\{k''\}$ for $1\le k<n$, where $(k',k'')=(k+1,k+2)$ if
$k$ is odd, $(k+2,k+1)$ if $k$ is even, except that
$U_{n-1}=\emptyset$ if $n$ is odd, $V_{n-1}=\emptyset$ if
$n$ is even.  Also $U_0=\{1\}$.  Therefore $\\{prev}[k]=0$ for
all $k$, and $\\{gen}[k](\l)$ reduces to the coroutine
$\\{nudge}[k]$ of Section~3.

\section 6. Launching. Ever since 1968, Section 1.4.2 of {\sl
The Art of Computer Programming\/} [\Ki] has contained the
following remark:  ``Initialization of coroutines tends to
be a little tricky, although not really difficult.''  Perhaps
that statement needs to be amended, from the standpoint of
the coroutines considered here.  We need to decide at which label
each coroutine $\\{gen}[k]$
should begin execution when it is first invoked: awake0,
asleep1, awake1, or asleep0.  And our discussion in
Sections~3 and~4 shows that we also need to choose the
initial setting of $a_1\ldots a_n$ very carefully.

Let's consider the initialization of $a_1\ldots a_n$ first.
The reflected Gray path mechanism that we use to construct
the paths $P_k$ and $Q_k$, as explained in Section~4,
complements some of the bits.  If, for example,
$U_k=\{u_1,u_2,\ldots,u_m\}$, where 
$u_1<u_2<\cdots<u_m$, path $P_k$ will contain
$n_{u_1}n_{u_2}\ldots n_{u_m}$ bit patterns, and the value
of bit $a_{u_i}$ at the end of $P_k$ will equal the value it
had at the beginning if and only if $n_{u_1}n_{u_2}\ldots
n_{u_{i-1}}$ is even.  The reason is that subpath $G_{u_i}$ is
traversed $n_{u_1}n_{u_2}\ldots n_{u_{i-1}}$  times, alternately
forward and backward.

In general, let
$$\delta_{jk}=\prod_{\scriptstyle u<j_{\mathstrut}\atop
                     \scriptstyle u\in U_k}
  n_u,{\rm\ if\ } j\in U_k; \qquad
  \delta_{jk}=\prod_{\scriptstyle v<j_{\mathstrut}\atop
                     \scriptstyle v\in V_k}
  n_v, {\rm\ if\ } j\in V_k.$$
Let $\alpha_{jk}$ and $\omega_{jk}$ be the initial and final
values of bit~$a_j$ in the Gray path~$G_k$ for spider~$k$,
and let $\tau_{jk}$ be the value of $a_j$ at the transition
point (the end of $P_k$ and the beginning of $Q_k$).  Then
$\alpha_{kk}=0$, $\omega_{kk}=1$, and the construction in
Section~4 defines the values of 
$\alpha_{ik}, \tau_{ik},$ and $\omega_{ik}$ for
$k<i\le\scope(k)$ as follows:  
Suppose $i$ belongs to spider~$j$, where $j$~is a child
of~$k$.  

\smallskip
\item{$\bullet$} If $j$ is positive, so that $j$ is a
principal element of $U_k$, we have $\tau_{ik}=\omega_{ij}$,
since $P_k$ ends with $a_j=1$.  Also
$\alpha_{ik}=\omega_{ij}$ if $\delta_{jk}$ is even,
$\alpha_{ik}=\alpha_{ij}$ if $\delta_{jk}$ is odd.  If
$k\to^*i$ we have $\omega_{ik}=1$; otherwise $i$ belongs to
spider $j'$, where $j'$ is a nonprincipal element of $V_k$.
In the latter case $\omega_{ik}=\alpha_{ij'}$ if 
$\omega_{j'j}+\delta_{j'k}$ is even, otherwise
$\omega_{ik}=\omega_{ij'}$.  (This follows because
$\omega_{j'j}=\tau_{j'k}$ and $\omega_{j'k}=(\tau_{j'k}+\delta_{j'k})\bmod
2$.)
\smallskip
\item{$\bullet$} If $j$ is negative, so that $j$ is a
principal element of $V_k$, we have $\tau_{ik}=\alpha_{ij}$,
since $Q_k$ begins with $a_j=0$.  Also
$\omega_{ik}=\alpha_{ij}$ if $\delta_{jk}$ is even,
$\omega_{ik}=\omega_{ij}$ if $\delta_{jk}$ is odd.  If
$i\to^*k$ we have $\alpha_{ik}=0@$; otherwise $i$ belongs to
spider $j'$, where $j'$ is a nonprincipal element of~$U_k$.
In the latter case $\alpha_{ik}=\alpha_{ij'}$ if 
$\alpha_{j'j}+\delta_{j'k}$ is even, otherwise
$a_{ik}=\omega_{ij'}$.

\smallskip\noindent
For example, when the digraph is the spider of Section~4,
these formulas yield
$$\vbox{\halign{#\quad&\hfil#\hfil&\quad$\hfil#={}$&#\hfil&&\qquad#\hfil\cr
$k$&$n_k$&\multispan2\hfil Initial bits $\alpha_{jk}$\hfil&
 \omit Transition bits $\tau_{jk}$&\omit\quad Final bits $\omega_{jk}$\cr
9  &  2  &         a_9&0          &           $*$      &      1            \cr
8  &  3  &       a_8a_9&00        &           $*1$     &      11           \cr
7  &  2  &         a_7&0          &           $*$      &      1            \cr
6  &  3  &       a_6a_7&00        &           $*0$     &      11           \cr
5  &  2  &         a_5&0          &           $*$      &      1            \cr
4  &  2  &         a_4&0          &           $*$      &      1            \cr
3  &  3  &       a_3a_4&00        &           $*0$      &      11           \cr
2  &  8  &    a_2a_3a_4a_5&0000   &          $*111$    &     1101          \cr
1  & 60  & a_1a_2\ldots a_9&000001100&    $*11011100$  &   111111100       \cr
}}$$

Suppose $j$ is a negative child of $k$.  If $n_u$ is odd for
all elements $u$ of $U_k$ that are less than~$j$, then
$\delta_{ij}+\delta_{ik}$ is even for all $i\in U_j$, and it
follows that $a_{ik}=\tau_{ij}$ for $j<i\le\scope(j)$.
(If $i$ is in spider~$j'$, where $j'\in U_j\subseteq U_k$, then
$\alpha_{ik}$ is $\alpha_{ij'}$ or $\omega_{ij'}$ according as
$\alpha_{j'j}+\delta_{j'k}$ is even or odd, and
$\tau_{ij}$ is $\alpha_{ij'}$ or $\omega_{ij'}$ according as
$\alpha_{j'j}+\delta_{j'j}$ is even or odd; and we have
$\delta_{j'k}\equiv\delta_{j'j}$ mod~2.)
On the other hand, if $n_u$ is even for some $u\in U_k$ with
$u<j$, then $\delta_{ik}$ is even for all $i\in U_j$, and we
have 
$\alpha_{ik}=\alpha_{ij}$ for $j<i\le\scope(j)$.  This
observation makes it possible to compute the initial bits
$a_1\ldots a_n$ in $O(n)$ steps (see [\Spiders]).

The special nature of vertex~0 suggests that we define $\delta_{j0}=1$
for $1\le j\le n$,
because we use path~$P_0$ but not~$Q_0$. This convention makes each
component of the digraph essentially independent. (Otherwise, for example,
the initial setting of $a_1\ldots a_n$ would be $01\ldots1$ in the trivial
``\\{poke}'' case when the digraph has no arcs.)

Once we know the initial bits, we start $\\{gen}[k]$ at label awake0
if $a_k=0$, at label awake1 if $a_k=1$.

\goodbreak
\section 7. Optimization. The coroutines
$\\{gen}[1]$, \dots, $\\{gen}[n]$ solve the general
spider-squishing problem, but they might not run very fast.
For example, the $\\{bump}$ routine in Section~2 takes an average of about
$n/2$ steps to decide which bit should be
changed.  We would much prefer to use only a bounded amount
of time per bit change, on the average, and this goal turns out
to be achievable if we optimize the coroutine
implementation.

A brute-force implementation of the \\{gen} coroutines, using only standard
features of Algol, can readily be written down based on an explicit stack and
a switch declaration:
\beginprogram
\quad Boolean \\{val}; \ comment \rm the current value being returned;\cr
\quad integer array $\\{stack}[0:2*n]$; \ comment \rm saved values of $k$
 and $l$;\cr
\quad integer $k$, $l$, $s$; \ comment \rm the current coroutine,
 parameter, and stack height;\cr
\quad switch $\\{sw}:=\rm p1, p2, p3, p4, p5, p6, p7, p8, p9, p10, p11$;\cr
\quad integer array $\\{pos}[0:n]$; \ comment \rm coroutine positions;\cr
\noalign{\medskip}
\quad $\langle\,$\rm Initialize everything$\,\rangle$;\cr
\quad \label p1: if $\\{maxu}[k]\ne0$ then begin\cr
\qquad $\\{invoke}(\\{maxu}[k],k,2)$;\cr
\qquad \label p2: if \\{val} then $\\{ret}(1)$;\cr
\qquad end;\cr
\quad $a[k]:=1$;  $\\{val}:=\\{true}$;  $\\{ret}(3)$;\cr
\quad \label p3: if $\\{maxv}[k]\ne0$ then begin\cr
\qquad $\\{invoke}(\\{maxv}[k],k,4)$;\cr
\qquad \label p4: if \\{val} then $\\{ret}(3)$;\cr
\qquad end;\cr
\quad if $\\{prev}[k]>l$ then begin\cr
\qquad $\\{invoke}(\\{prev}[k],l,5)$;\cr
\qquad \label p5: $\\{ret}(6)$;\cr
\qquad end\cr
\quad else begin $\\{val}:=\\{false}$;  $\\{ret}(6)$; end;\cr
\quad \label p6: if $\\{maxv}[k]\ne0$ then begin\cr
\qquad $\\{invoke}(\\{maxv}[k],k,7)$;\cr
\qquad \label p7: if \\{val} then $\\{ret}(6)$;\cr
\qquad end;\cr
\quad $a[k]:=0$;  $\\{val}:=\\{true}$;  $\\{ret}(8)$;\cr
\quad \label p8: if $\\{maxu}[k]\ne0$ then begin\cr
\qquad $\\{invoke}(\\{maxu}[k],k,9)$;\cr
\qquad \label p9: if \\{val} then $\\{ret}(8)$;\cr
\qquad end;\cr
\quad if $\\{prev}[k]>l$ then begin\cr
\qquad $\\{invoke}(\\{prev}[k],l,10)$;\cr
\qquad \label p10: $\\{ret}(1)$;\cr
\qquad end\cr
\quad else begin $\\{val}:=\\{false}$; $\\{ret}(1)$; end;\cr
\quad \label p11: $\langle\,$\rm Actions of the driver program when
$k=0\,\rangle$;\cr
\endprogram
Here $\\{invoke}(\\{newk},\\{newl},j)$ is an abbreviation for
\beginprogram
$\\{pos}[k]:=j$; \ $\\{stack}[s]:=k$; \ $\\{stack}[s+1]:=l$; \
 $s:=s+2$;\cr
$k:=\\{newk}$; \ $l:=\\{newl}$; \ go to $\\{sw}[\\{pos}[k]]$\cr
\endprogram
and $\\{ret}(j)$ is an abbreviation for
\beginprogram
$\\{pos}[k]:=j$; \ $s:=s-2$;\cr
$l:=\\{stack}[s+1]$; \ $k:=\\{stack}[s]$; \ go to $\\{sw}[\\{pos}[k]]$\rm.\cr
\endprogram

We can streamline the brute-force implementation in several straightforward
ways. First we can use a well-known technique to simplify the ``tail
recursion'' that occurs when \\{invoke} is immediately followed by \\{ret}
(see [\Ksp, example 6a]): The statements `$\\{invoke}(\\{prev}[k],l,5)$;
\ p5:\ \\{ret}(6)' can, for example, be replaced by
$$\\{pos}[k]:=6;\ \ k:=\\{prev}[k];\ \ \hbox{\bf go to }\\{sw}[\\{pos}[k]].$$

An analogous simplification is possible for the constructions of the form
`{\bf while $A$ do return \\{true}}' that occur in $\\{gen}[k]$.
For example, we could set things up so that coroutine~$A$ removes {\it two\/}
pairs of items from the stack when it returns with $\\{val}=\\{true}$,
if we first set $\\{pos}[k]$ to the index of a label that follows the {\bf
while} statement. More generally, if coroutine~$A$ itself 
is also performing
such a {\bf while} statement, we could allow {\bf return} statements to
remove even more than two pairs of stack items at a time. Details are left to
the reader.

\section 8.  The active list. The $\\{gen}$ coroutines of
Section 5 perform $O(n)$ operations per bit change, as they
pass signals back and forth, because each coroutine carries
out at most two lines of its program.  This upper bound on
the running time cannot be substantially improved, in
general.  For example, the $\\{bump}$ coroutines of
Section~2 typically need to interrogate about \smash{${1\over2}n$}
trolls per step; and it can be shown that the $\\{nudge}$
coroutines of Section~3 typically involve action by about
$cn$ trolls per step, where $c=(5+\sqrt{5}@)/10\approx0.724$.
(See [\Ki, exercise 1.2.8--12].)

Using techniques like those of Section~7, however, the
$\\{gen}$ coroutines can always be transformed into a
procedure that performs only $O(1)$ operations per bit
change, amortized over all the changes.  A formal derivation
of such a transformation is beyond the scope of the present
paper, but we will be able to envision it by considering an
informal description of the algorithm that results.

The key idea is the concept of an {\it active list}, which
encapsulates a given stage of the computation.  The active
list is a sequence of nodes that are either awake or asleep.
If $j$ is a positive child of~$k$, node $j$ is in the active
list if and only if $k=0$ or $a_k=0@$; if $j$ is a negative
child of $k$, it is in the active list if and only if
$a_k=1$.

Examples of the active list in special cases have appeared
in the tables illustrating $\\{bump}$ in Section~2 and
$\\{nudge}$ in Section~3.  Readers who wish to review those
examples will find that the numbers listed there do indeed
satisfy these criteria.  Furthermore, a node number has been
underlined when that node is asleep; bit $a_j$ has been
underlined if and only if $j$ is asleep and in the active
list.

Initially $a_1\ldots a_n$ is set to its starting pattern as
defined in Section~6, and all elements of the corresponding
active list are awake.  To get to the next bit pattern, we
perform the following actions:

\smallskip
\item{1)}  Let $k$ be the largest nonsleeping node on the
active list, and wake up all nodes that are larger.  (If all
elements of the active list are asleep, they all wake up and
no bit change is made; this case corresponds to $\\{gen}
[\\{maxu}[0]](0)$ returning $\\{false}$.)
\item{2)}  If $a_k=0$, set $a_k$ to~1, delete $k$'s
positive children from the active list, and insert $k$'s
negative children.  Otherwise set $a_k$ to~0, insert the
positive children, and delete the negative ones.  (Newly
inserted nodes are awake.)
\item{3)}  Put node $k$ to sleep.

\smallskip\noindent
Again the reader will find that the $\\{bump}$ and
$\\{nudge}$ examples adhere to this discipline.

If we maintain the active list in order of its nodes, the
amortized cost of these three operations is $O(1)$, because
we can charge the cost of inserting, deleting, and awakening
node~$k$ to the time when bit $a_k$ changes.  Steps (1) and
(2) might occasionally need to do a lot of work, but this
argument proves that such difficult transitions must be
rare.

Let's consider the spider of Section~4 one last time.  The
60 bit patterns that satisfy its constraints are generated
by starting with $a_1\ldots a_9=000001100$, as we observed
in Section~6, and the Gray path $G_1$ begins as follows
according to the active list protocol:
$$\vbox{\halign{#\qquad&#\cr
000001100&1235679\cr
00000110\b1&123567\b9\cr
000001\b001&12356\b79\cr
000001\b00\b0&12356\b7\b9\cr
00000\b0000&1235\b69\cr
00000\b000\b1&1235\b6\b9\cr
0000\b10001&123\b569\cr
0000\b1000\b0&123\b56\b9\cr
0000\b1\b1000&123\b5\b679\cr}}$$
(Notice how node 7 becomes temporarily inactive when $a_6$
becomes 0.)  The most dramatic change will occur after the
first $n_2n_6n_9=48$ patterns, when bit $a_1$ changes as we
proceed from path $P_1$ to path $Q_1$:
$$\vbox{\halign{#\qquad&#\cr
0\b11\b01\b1\b10\b0&1\b2\b4\b6\b7\b9\cr
\b111011100&\b14789\cr}}$$
(The positive children 2 and 6 have been replaced by the negative
child~8.)
Finally, after all 60 patterns have been generated, the
active list will be $\b1\b4\b7\b8\b9$ and $a_1\ldots a_9$
will be $\b111\b111\b1\b0\b0$.  All active nodes will be
napping, but when we wake them up they will be ready to
regenerate the 60 patterns in reverse order.

It should be clear from these examples, and from a careful
examination of the $\\{gen}$ coroutines, that steps (1), (2),
and (3) faithfully implement those coroutines in an
efficient iterative manner.

\section 9. Additional optimizations. The algorithm of Section
8 can often be streamlined further.  For example, if $j$ and
$j'$ are consecutive positive children of $k$ and if $V_j$
is empty, then $j$ and $j'$ will be adjacent in the active
list whenever they are inserted or deleted.  We can
therefore insert or delete an entire family en masse, in the
special case that all nodes are positive, if the active list
is doubly linked.  This important special case was first
considered by Koda and Ruskey [\KR]; see also [\Kiv,
Algorithm 7.2.1.1K].

Further tricks can in fact be employed to make the active
list algorithm entirely 
{\it loopless}, in the sense that $O(1)$ operations are
performed between successive bit changes in {\it all\/} cases\dash---not
only in an average, amortized sense.  One idea, used
by Koda and Ruskey in the special case just mentioned, is to
use ``focus pointers'' to identify the largest nonsleeping
node (see [\Ehrlich] and [\Kiv, Algorithm 7.2.1.1L]).
Another idea, which appears to be necessary when both
positive and negative nodes appear in a complex family, is to
perform lazy updates to the active list, changing links only
gradually but before they are actually needed.  Such a
loopless implementation, which moreover needs only $O(n)$
steps to initialize all the data structures, is described
fully in [\Spiders].  It does not necessarily run faster
than a more straightforward amortized $O(1)$ algorithm, from
the standpoint of total time on a sequential computer; but
it does prove that a strong performance guarantee is
achievable, given any totally acyclic digraph.

\section 10.  Conclusions and acknowledgements.  We have seen
that a systematic use of cooperating coroutines leads to a
generalized Gray code for generating all bit patterns that
satisfy the ordering constraints of any totally acyclic
digraph.  Furthermore those coroutines can be implemented
efficiently, yielding an algorithm that is faster than
previously known methods for that problem.  Indeed, the
algorithm is optimum, in the sense that its running time is
linear in the number of outputs.

Further work is clearly suggested in the heretofore
neglected area of coroutine transformation.  For example, we
have not discussed the implementation of coroutines such~as 
\beginprogram
\quad Boolean coroutine $\\{copoke}[k]$;\cr
\qquad while $\\{true}$ do begin\cr
\quad\qquad if $k<n$ then while $\\{copoke}[k+1]$ do return $\\{true}$;\cr
\quad\qquad $a[k]:=1-a[k]$; return $\\{true}$;\cr
\quad\qquad if $k<n$ then while $\\{copoke}[k+1]$ do return $\\{true}$;\cr
\quad\qquad return $\\{false}$;\cr
\quad\qquad end\rm.\cr
\endprogram
These coroutines, which are to be driven by repeatedly
calling $\\{copoke}[1]$, generate Gray binary code, so their effect is
identical to repeated calls on the coroutine $\\{poke}[n]$ in Section~2.
But \\{copoke} is much less efficient, since \\{copoke}[1] always
invokes \\{copoke}[2], \dots, $\\{copoke}[n]$ before returning a result.
Although these \\{copoke} coroutines look superficially similar to
$\\{gen}$, they are not actually a special case of that
construction.  A rather large family of coroutine
optimizations seems to be waiting to be discovered and to be
treated formally.

Another important open problem is to discover a method that
generates the bit patterns corresponding to an {\it
arbitrary\/} acyclic digraph, with an amortized cost of only
$O(1)$ per pattern.  The best currently known bound is
$O(\log n)$, due to M.~B.\ Squire [\Squire]; see also
[\Ruskey, Section 4.11.2]. There is always a listing of the
relevant bit patterns in which at most two bits change from one
pattern to the next~[\PR, Corollary~1].

The first author thanks Ole-Johan Dahl for fruitful
collaboration at the University of Oslo during 1972--1973 and
at Stanford University during 1977--1978; also for sharing
profound insights into the science of programming and for
countless hours of delightful four-hands piano music over a period of more
than 30~years.  The
second author thanks Malcolm Smith and Gang (Kenny) Li for
their help in devising early versions of algorithms
for spider-squishing during 1991 and 1995, respectively.
Both authors are grateful to Stein Krogdahl and to an anonymous referee,
whose comments on a previous draft of this paper have led to substantial
improvements.

\vfill\eject
\centerline{\bf References}

\bigskip\noindent
[\Clint] M. Clint, ``Program proving: Coroutines,'' {\sl Acta
Informatica\/ \bf2} (1977), 50--63.

\bigskip\noindent
[\Conway] Melvin E. Conway, ``Design of a separable transition-diagram
compiler,'' {\sl Communications of the ACM\/ \bf6} (1963), 396--408.

\bigskip\noindent
[\Sbase] Ole-Johan Dahl, Bj{\o}rn Myhrhaug, and Kristen Nygaard,
{\sl SIMULA-67 Common Base Language}, Publication S-2
(Oslo:\ Norwegian Computing Center, 1968), 141 pages. Revised
edition, Publication S-22 (1970), 145 pages. Third revised edition,
Report number 725 (1982), 127 pages.

\bigskip\noindent
[\SP] Ole-Johan Dahl and C. A. R. Hoare, ``Hierarchical program
structures,'' in {\sl Structured Programming\/} (Academic Press,
1972), 175--220.

\bigskip\noindent
[\Dsyn] Ole-Johan Dahl, {\sl Syntaks og Semantikk i
Programmeringsspr{\aa}k\/} (Lund:\ Student\-littera\-tur, 1972), 103 pages.

\bigskip\noindent
[\Dsemi] Ole-Johan Dahl, ``An approach to correctness proofs of
semicoroutines,'' Research Report in Informatics, Number 13 (Blindern,
Norway:\ University of Oslo, 1977), 20 pages.

\bigskip\noindent
[\Ehrlich] Gideon Ehrlich, ``Loopless algorithms for
generating permutations, combinations and other
combinatorial configurations,'' {\sl Journal of the
Association for Computing Machinery\/ \bf 20} (1973),
500--513.

\bigskip\noindent
[\Floyd] Robert W. Floyd, ``The syntax of programming languages --- A
survey,'' {\sl IEEE Transactions on Electronic Computers\/ \bf EC-13}
(1964), 346--353.

\bigskip\noindent
[\Ki] Donald E.\ Knuth, {\sl Fundamental Algorithms}, Volume
1 of {\sl the Art of Computer Programming\/} (Reading,
Massachusetts: Addison--Wesley, 1968).  Third edition, 1997.

\bigskip\noindent
[\KO] Donald E.\ Knuth, {\sl Selected Topics in Computer
Science, Part II}, Lecture Note Series, Number 2
(Blindern, Norway: University of Oslo, Institute of
Mathematics, August 1973). See page 3 of the notes entitled
``Generation of combinatorial patterns: Gray codes.''

\bigskip\noindent
[\Ksp] Donald E. Knuth, ``Structured programming with {\bf go to}
statements,'' {\sl Computing Surveys\/ \bf 6} (December 1974),
261--301. Reprinted with revisions as
Chapter~2 of {\sl Literate Programming\/} (Stanford, California:\
Center for the Study of Language and Information, 1992).

\bigskip\noindent
[\Kiv] Donald E.\ Knuth, ``Generating all $n$-tuples,''
Section 7.2.1.1 of {\sl The Art of Computer Programming},
Volume 4 (Addison--Wesley), in preparation. Preliminary
excerpts of this material are available at 
{\tt http://www-cs-faculty.stanford.\allowbreak
edu/\char`~knuth/news01.html}.

\bigskip\noindent
[\Spiders] Donald E.\ Knuth, {\mc SPIDERS}, a program downloadable
from the website\hfill\break\null\qquad
{\tt http://www-cs-faculty.stanford.edu/\char`~knuth/programs.html}.

\bigskip\noindent
[\KR] Yasunori Koda and Frank Ruskey, ``A Gray code for the
ideals of a forest poset,'' 
{\sl Journal of Algorithms\/ \bf 15} (1993), 324--340.

\bigskip\noindent
[\PR] Gara Pruesse and Frank Ruskey, ``Gray codes from antimatroids,''
{\sl Order\/ \bf10} (1993), 239--252.

\bigskip\noindent
[\Ruskey] Frank Ruskey, {\sl Combinatorial Generation\/}
[preliminary working draft].  Department of Computer
Science, University of Victoria, Victoria B.C., Canada
(1996).

\bigskip\noindent
[\Squire] Matthew Blaze Squire, {\sl Gray Codes and
Efficient Generation of Combinatorial Structures}. Ph.D.\
dissertation, North Carolina State University (1995), $\rm
x+145$ pages.

\bigskip\noindent
[\Steiner] George Steiner, ``An algorithm to generate the
ideals of a partial order,'' {\sl Operations Research
Letters\/ \bf 5} (1986), 317--320. 

\bigskip\noindent
[\VM] Leonard I. Vanek and Rudolf Marty, ``Hierarchical coroutines: A method
for improved program structure,'' {\sl Proceedings of the 4th International
Conference on Software Engineering\/} (Munich, 1979), 274--285.

\bigskip\noindent
[\WD] Arne Wang and Ole-Johan Dahl, ``Coroutine sequencing in a
block-structured environment,'' {\sl BIT\/ \bf11} (1971), 425--449.

\bye